\documentclass[fleqn]{2020SCGE}
\usepackage[T1]{fontenc}
\usepackage{tgschola}

\newcommand{\mnras}{Mon. Not. R. Astron. Soc}
\newcommand{\apjl}{The Astrophysical Journal Letters}
\newcommand{\apjs}{The Astrophysical Journal Supplement Series}
\newcommand{\pasa}{Publ. Astron. Soc. Aust}
\newcommand{\asa}{Astron. Astrophys}
\newcommand{\aj}{Astron. J}

\begin{document}

\ensubject{subject}

\ArticleType{Invited Review}
\Year{2022}
\Month{ }
\Vol{ }
\No{ }
\DOI{??}
\ArtNo{000000}
\ReceiveDate{ }
\AcceptDate{ }

\title{Status and progress of China SKA Regional Centre prototype}{China SRC prototype}


\author[1,2]{Tao An}{antao@shao.ac.cn}
\author[1]{Xiaocong Wu}{}
\author[1]{Baoqiang Lao}{}
\author[1]{Shaoguang Guo}{}
\author[1]{Zhijun Xu}{}
\author[1]{\\ Weijia Lv}{}
\author[1]{Yingkang Zhang}{}
\author[1]{Zhongli Zhang}{}

\AuthorMark{An T}

\AuthorCitation{An T, et al}


\address[1]{Shanghai Astronomical Observatory, Chinese Academy of Sciences, Shanghai, 200030, China}
\address[2]{Peng Cheng Laboratory, Shenzhen, 518066, China}


\abstract{The Square Kilometre Array (SKA) project consists of delivering two largest radio telescope arrays being built by the SKA Observatory (SKAO), which is an intergovernmental organization bringing together nations from around the world with China being one of the major member countries. The computing resources needed to process, distribute, curate and use the vast amount of data that will be generated by the SKA telescopes are too large for the SKAO to manage on its own. To address this challenge, the SKAO is working with the international community to create a shared, distributed data, computing and networking capability called the SKA Regional Centre Alliance. In this model, the SKAO will be supported by a global network of SKA Regional Centres (SRCs) distributed around the world in its member countries to build an end-to-end science data system that will provide astronomers with high-quality science products. SRCs undertake deep processing, scientific analysis, and long-term storage of the SKA data, as well as user support. China has been actively participating in and promoting the construction of SRCs. This paper introduces the international cooperation and ongoing prototyping of the global SRC network, the basis for the construction of the China SRC and describes in detail the progress of the China SRC prototype. The paper also presents examples of scientific applications of SKA precursor and pathfinder telescopes performed using resources from the China SRC prototype. Finally, the future prospects of the China SRC are presented.}



\maketitle


\begin{multicols}{2}
\section{Introduction}\label{sec:intro}

The Square Kilometre Array (hereinafter referred to as SKA) \footnote{\url{https://www.skao.int}}, an international large science project, is the world's largest radio telescope under construction by a joint effort of more than a dozen countries around the world. It is named after its total collection area of one square kilometer when fully completed.
\Authorfootnote
The SKA is a synthesis aperture array consisting of a large number of small-diameter radio antennas stretching over thousands of kilometers. Its extraordinary capabilities \cite{2009IEEEP..97.1482D}, such as large field of view, high sensitivity, high resolution, wide frequency coverage and fast surveys, provide significant opportunities for humans to explore the mysteries of the Universe. SKA aims to make profound discoveries in a number of frontier scientific areas, including the formation of the first stars and the first galaxies, the formation and evolution of galaxies, the nature of dark energy, the origin of the cosmic magnetic fields, the nature of gravity, and extraterrestrial civilizations \cite{wu2019,skasw2015}. 

The SKA Observatory (SKAO) consists of three components: a global headquarter located at the Jodrell Bank Observatory in the United Kingdom, a low-frequency telescope (SKA-Low) in Australia and a mid-frequency telescope (SKA-Mid) in South Africa and eight southern African countries. The construction of SKA is divided into two phases: Phase 1 (SKA1, 2021--2029, 10\% of the whole scale), Phase 2 (SKA2, after 2030, the full scale). 
In SKA1, the telescope elements are concentrated in the core array with a maximum baseline of 65 km for SKA1-Low and 100 km for SKA1-Mid, respectively \cite{SKA1design}.

\begin{figure*}
    \centering
    \includegraphics[width=0.9\textwidth]{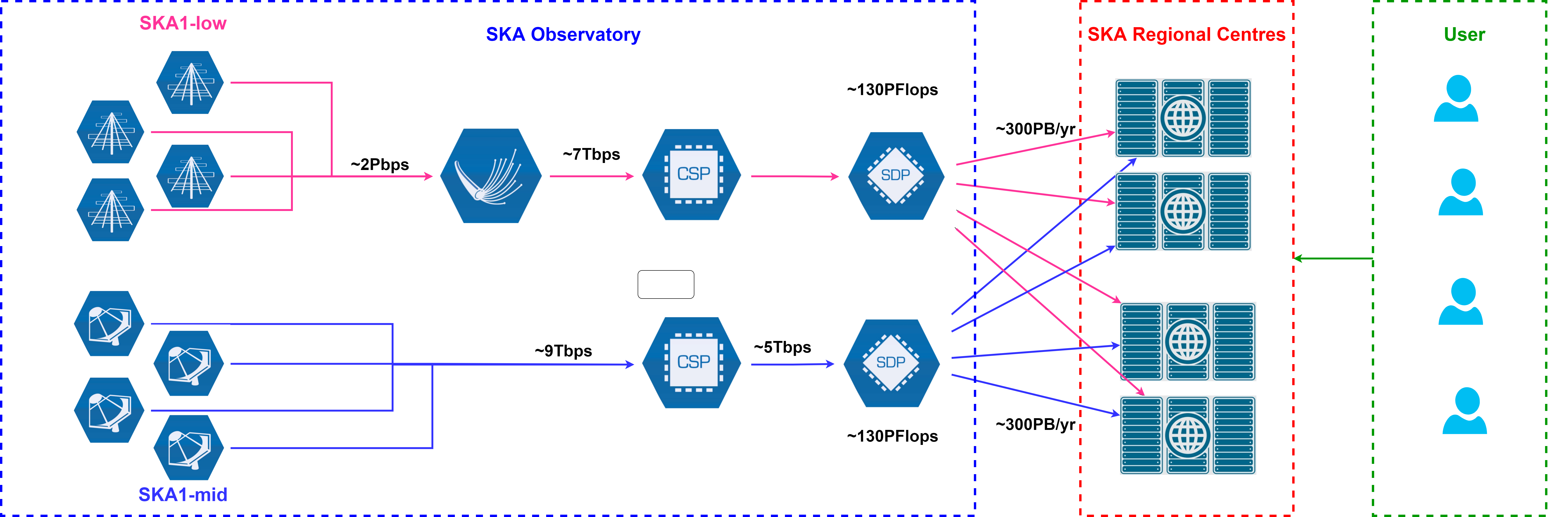}
    \caption{The sketchmap of the SKA data flow showing the data rate at each major stage. Copyright\copyright SKA Observatory. Reproduced with permission from the SKA Observatory.}
    \label{fig:SKAdataflow}
\end{figure*}

China is a major member of the SKA and has actively participated in the entire SKA construction process starting from the early site selection. In June 2021, China signed the SKA Observatory Convention, bringing China's participation in the SKA to a new stage \cite{wang2022}. At the time of writing, eight countries, including China, have acceded the SKA Observatory Convention. 

As a revolutionary telescope in the history of radio astronomy, the SKA is expected to achieve outstanding scientific success, depending not only on the performance of its telescope, but also on the ability and capability of astronomers to process and analyze SKA data \cite{wu2019}. The amount of SKA science data is unprecedented in astronomy. 
Figure \ref{fig:SKAdataflow} shows a sketch of the SKA data flow including several major stages.
After beamforming, SKA1 data are fed into the science data processors (SDPs, the supercomputers of the SKAO to process the observational data) for pre-processing at a rate of 5--7 terabits per second (Tb/s, 1 Tb = $10^{12}$ bits) \cite{SKA1design}. After the initial calibration and data quality check in the SDPs, $\sim$700 petabytes (PB, 1 PB = $10^{15}$ Bytes) of scientific data will be generated annually \cite{SRCrequirement,an2019bigdata}. 

The resources required to fully process, distribute, curate, and utilize SKA data are beyond the budget of the SKAO. 
The SKAO and the international SKA science community are jointly planning multiple regional science and data centers (SKA Regional Centres, or SRCs). These SRCs are intended to establish globally shared capacities for distributed data, computing, and networking to foster international collaboration and support a broad range of SKA science cases \cite{SRCWP2020} (Figure \ref{fig:SKAdataflow}).

At a fixed cost, the primary task of the SDPs is to process SKA observational data in real time, and the SDPs have no excess capacity for deep processing of the data. However, the pre-processed data output from SDPs cannot be directly used for scientific research, so a large amount of complex computing tasks need to be supplemented by additional capabilities from the SRCs. 
Moreover, by its conceptual design, the SDP supports single-directional workflows and has only a few hours of buffer space, requiring that the buffer to be emptied in a timely manner to ensure subsequent data injection. 
Many discoveries are generated through multiple iterations of data processing and deep data mining. For example, the first fast radio burst was discovered from the analysis of archival radio data \cite{2007Sci...318..777L}.
Whereas, general users do not have read access to the data in the SDP. 
A large amount of scientific data will be stored in the SRCs for long periods for scientific users to use. 
SKA has a large and globally distributed user community, and the computing tasks used by astronomers for different scientific projects may rely on different computing architectures and software environments, making it impossible for a single data center to provide flexible resource allocation and  convenient technical support to meet differentiated needs. 
Therefore, SRCs not only share the overall pressure of the SKAO data processing and curation, but also undertake user support worldwide to maximize the differentiated needs of users for the maximum benefits of the SKA scientific community.

\section{Progress and status of the global SRC network}

The complexity of the infrastructure capable of handling the expected volume of SKA data far exceeds the current capabilities and technologies of the radio astronomy community. 
Over the last decade or so, 
some research projects have been imitated for SRC conceptual design and prototype development, such as Advanced European Network of E-infrastructures for Astronomy with the SKA (AENEAS) \footnote{\url{https://www.aeneas2020.eu/}} and Exascale Research Infrastructure for Data in Asia-Pacific Astronomy Using The SKA (ERIDANUS) \footnote{\url{https://eridanus.net.au}}. 

AENEAS is 
funded by the European Commission's Horizon 2020 programme to develop the concept and design of a distributed, federated European Science Data Centre (ESDC) for the needs of SRCs. AENEAS investigated the innovative solutions for networking, access, processing and storage needed for SKA data processing, providing important references for the architecture and technical solutions for the SRCs.

ERIDANUS 
focuses on the deployment of data-intensive research infrastructure prototypes and middleware in Australia and China SRCs, with the aim of addressing SKA1-level data processing challenges. The main players in the project are International Centre for Radio Astronomy Research (ICRAR, Australia), Commonwealth Scientific and Industrial Research Organisation (CSIRO, Australia) and Shanghai Astronomical Observatory (SHAO, China). 

The main joint outcomes of ERIDANUS are: 
(1) the establishment of the networking and middleware between the astronomical community and industry (research network providers, supercomputing centers, commercial cloud providers). A representative user case is the transcontinental network test experiments that has been carried out between Shanghai and Perth supercomputer with a maximum bandwidth of 5 Gbps and 90\% bandwidth utilization \cite{Guo2022-network}. (2) the completion of the Australian SRC White Paper; 
(3) the organization of two Australia-China SKA Big Data Workshops in Shanghai in 2017 and 2018, which are helpful for testing the SRC key technologies; 
(4) a joint team from ICRAR, Oak Ridge National Laboratory (ORNL, US), and SHAO successfully simulated a large workflow at the SKA1 scale \cite{2020SciBu..65..337W,2020hpcn.conf...11W}, a milestone in the development of SKA data processing system (see Section~\ref{sec:CSRC-sci-4} for details). This outstanding achievement in high performance computing (HPC) applications was selected as one of six finalists for the prestigious Gordon Bell Prize \footnote{\url{https://www.icrar.org/gordon-bell-prize-finalist/}}. 
(5) The China-Australia team performed benchmark performance tests on the China SRC prototype.
Using the same number of nodes from the China SRC prototype and the then ASKAP data center to process the same ASKAP data showed that the China SRC prototype had better performance, as has been demonstrated in the 2019 SKA Engineering Meeting\footnote{\url{https://indico.skatelescope.org/event/551/contributions/6734/}}. 
The design concept of the China SRC prototype was adopted by other SRCs later. 

To further coordinate the global SRC collaborations,
in November 2018, the SKA Council approved the establishment of the SRC Steering Committee (SRCSC).
The SRCSC is responsible for developing the top-level planning, program design, governance model, funding requirements, prototype  testing, and construction plans for the SRCs. 

In the future, 
the collective contributions from each SRC will form the "SRC Network" (SRCNet) (Figure \ref{fig:SKAdataflow}), which together with the SKAO will create SKA data products. 
However, the complexity of the collaborative work between SKAO and  SRCNet has never been encountered by astronomers before.
Users have to request resources from both SKAO and SRCNet to complete a research project, where resources include not only SKAO's telescope time and SDP's supercomputing time, but also SRCNet's intercontinental network bandwidth for data transfer, data processing time, and data storage.
A necessary condition for the scientific success of the SKA is a coherent, persistent and powerful end-to-end science data processing system.

Therefore, the primary functions of the SRCNet include: establishing the smooth data logistics between SKAO and individual SRCs, ensuring user to access data stored at SRCs, deep processing and scientific analysis of SKA data, providing long-term storage and curation of SKA data, conducting research and development of SKA data-related technologies, and providing user support including education and outreach.

\begin{figure*}
    \centering
    \includegraphics[width=0.9\textwidth]{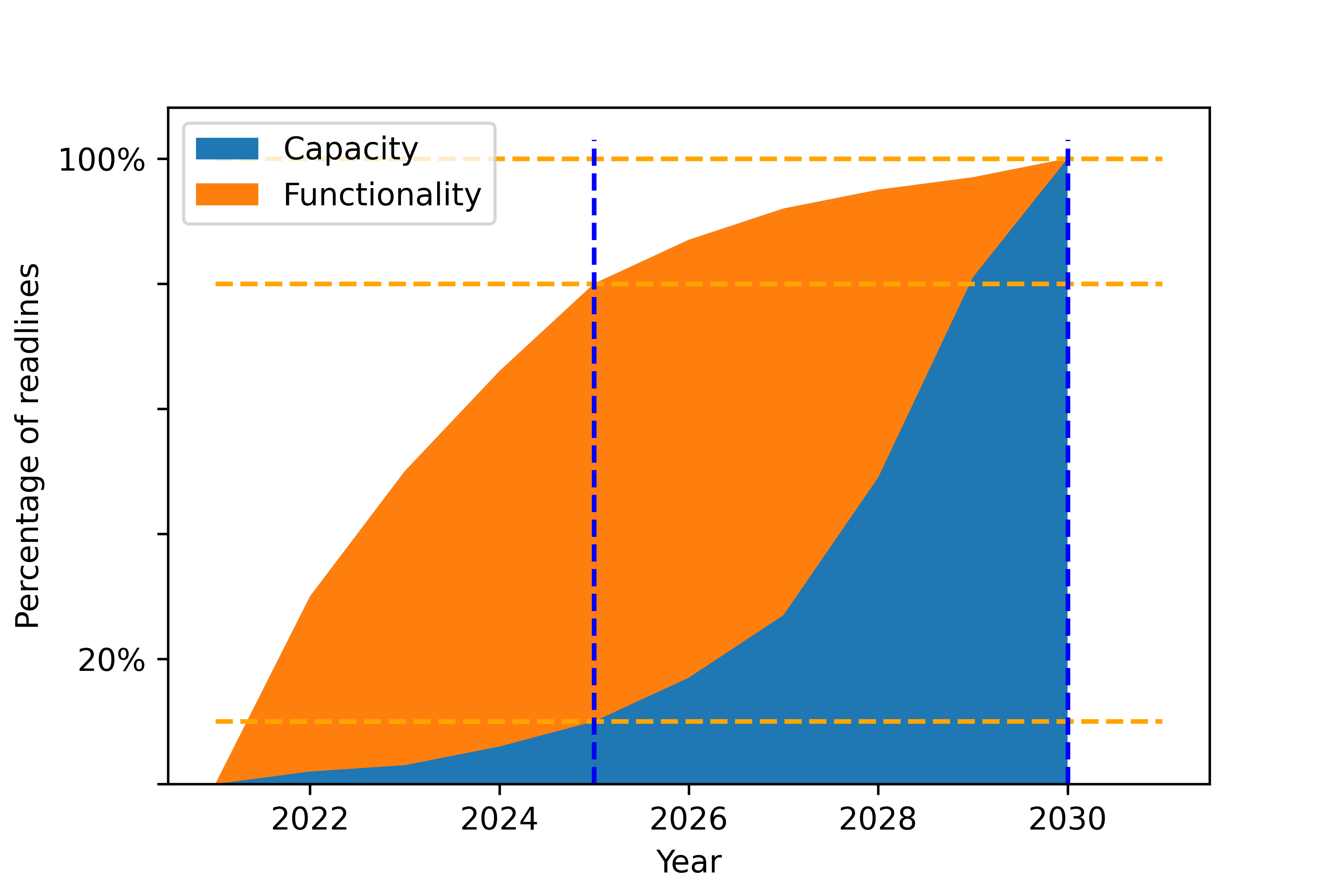}
    \caption{The timeline of the SKA Regional Centre. Reproduced from Ref. \cite{SRCWP2020}.}
    \label{fig:SRCscale}
\end{figure*}

\begin{figure*}
    \centering
    \includegraphics[width=0.9\textwidth]{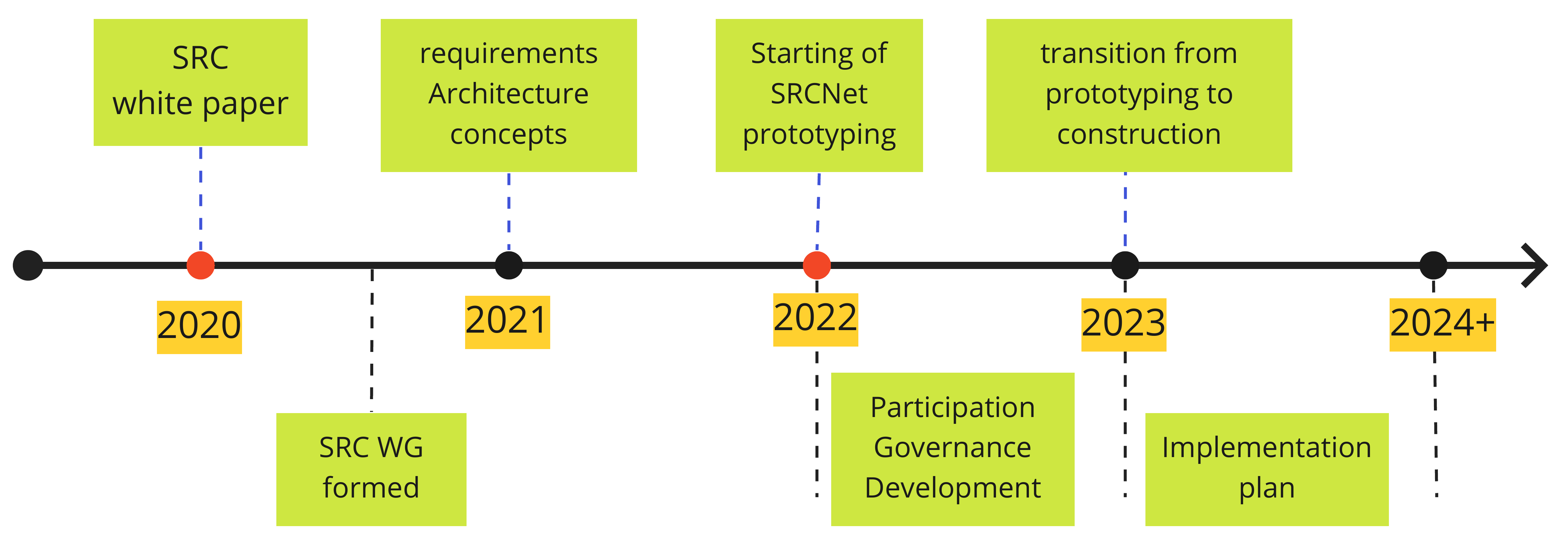}
    \caption{Roadmap of the SKA Regional Centre network in the first phase \cite{SRCWP2020}.}
    \label{fig:SRCNetPlan}
\end{figure*}

The past few years have been a critical period of not only the initial construction for the SKA Observatory but also the preparation for the SRCs (see the timeline of the SRC in the prototyping phase in Figure \ref{fig:SRCNetPlan}). In May 2020, the SRCSC completed the SRC White Paper (v1) \cite{SRCWP2020} which was approved by the SKA Council. 
The White Paper covers the positioning, functions, structure, scale, technical solutions and governance model of the future SRCs, and is the framework document for the overall planning of the SRCNet.

The SRC white paper focuses on three key issues \cite{SRCWP2020}: 
\begin{itemize}
\item Governance and operation models. 
Under the governance of the SKA Council, the SRCNet operates on two levels of governance and operations. 
The governance committee will be responsible for managing and developing the SRC resources to ensure they are persistent, coherent, and scalable. The operation group is responsible for day-to-day operations to ensure that the services and resources for data logistics are available to meet scientific needs.
\item Baseline functions and roadmap.
The SRCNet, when completed, will receive $\sim$700~PB data per year from the SKA Observatory, with a computing capacity of 22 Pflops and a network bandwidth of 100 Gbps for data transfer between multiple SRCs. This capability will be achieved in two phases (Figure \ref{fig:SRCscale}). 
From 2020 to 2025, requirements analysis, key technology development, prototype system construction and testing will be carried out to reach 10\% of the total capability and 80\% of the full functionalities by the end of 2024 (Figure \ref{fig:SRCNetPlan}). In 2025--2030, the SRCNet prototype will be scaled-up to the full capability and full functionality, and provides support to early science validation and operation of SKA1. 
\item National commitment. The basic principle is that SKA member countries invest in the SRC in the same proportion to their investment in the SKAO. The contributions by the stakeholders of the SRC are not limited to direct government investment and can take the form of in-kind contributions, cash and human resources.
\end{itemize}

\begin{figure*}
    \centering
    \includegraphics[width=0.8\textwidth]{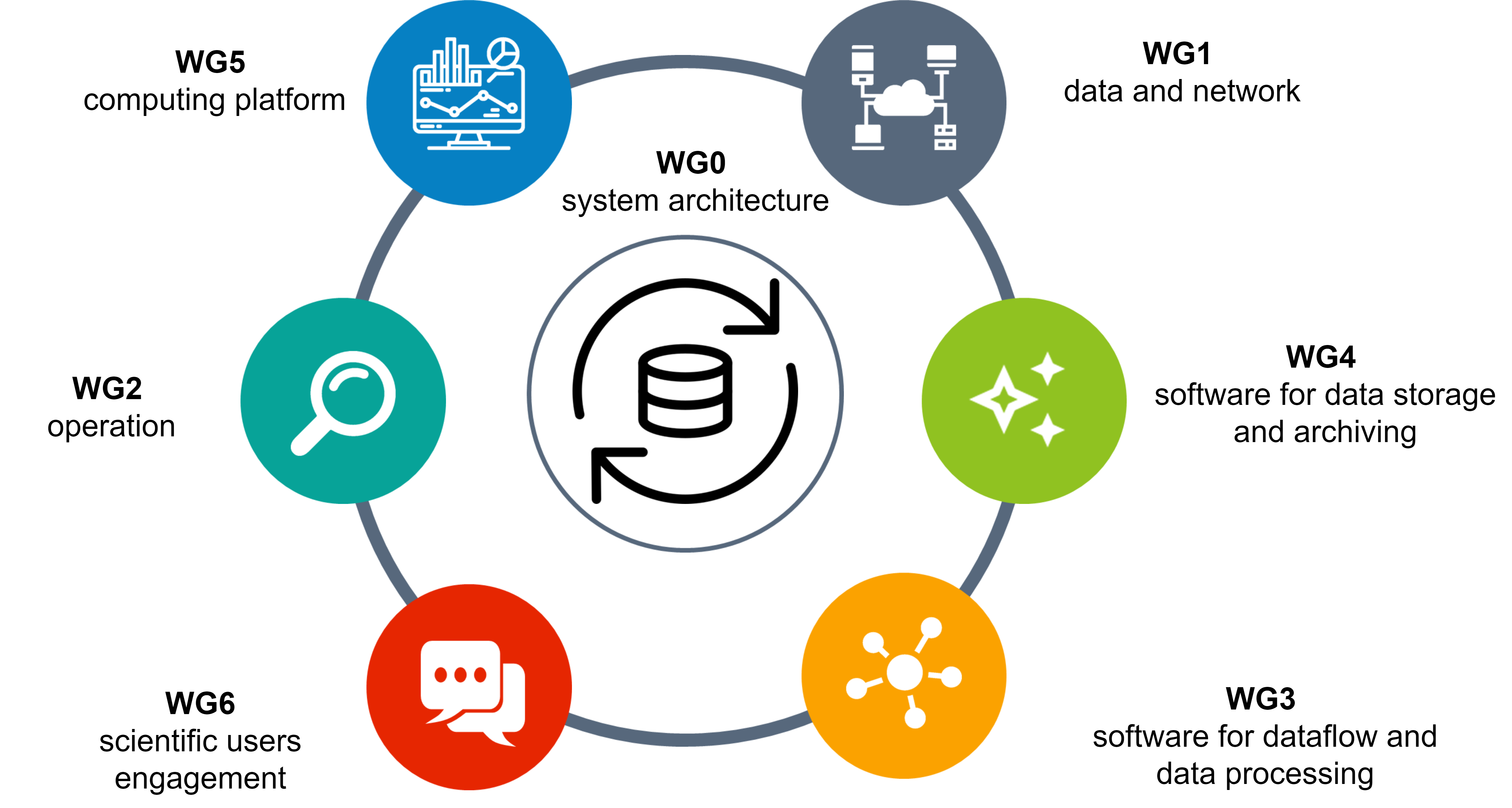}
    \caption{SKA regional centre working groups. WG0 (system architecture), WG1 (data and network), WG2 (operation), WG3 (software for dataflow and data processing), WG4 (software for data storage and archiving), WG5 (computing platform), and WG6 (scientific users engagement). WG6 interacts with scientific users and is a long-standing working group. Other working groups may play a role in different phases of SRC construction. Copyright\copyright SKA Observatory.}
    \label{fig:SRCwg}
\end{figure*}

Under the leadership of the SRC Steering Committee, seven SRC Working Groups (WG) have been established to carry out preparatory work (Figure \ref{fig:SRCwg}). 
Chinese scientists and engineers have already undertaken relevant tasks in the first six WGs.
In the early stages, the working groups are primarily supported by resources from individual SRC prototypes (including China SRC prototype)
and the SKAO.

From 2022 until mid-2023, the SKAO and SRCSC will focus on the SRCNet prototype testing (Figure \ref{fig:SRCNetPlan}). 
In the prototyping phase, the SKAO invited the China SRC team to lead one of the seven teams to play an important role in the SRCNet preparation phase. This team is composed of researchers from Australia, China, Japan and Korea SRCs, and is responsible for providing new SRC infrastructure or upgrading the existing SRC infrastructure to support the SRCNet prototyping. These prototyping teams work in the Scaled Agile Framework (SAFe) manner.
The SRC is being transformed from concept to reality.

In addition to the global cooperative SRCNet prototype, 
individual SRCs projects are being advanced in some SKA member countries (e.g., Australia, Canada, China, Netherlands, Portugal, Spain, UK).

Australia hosts the SKA-Low telescope, and has built two clusters dedicated to SKA precursor telescope (ASKAP and MWA, both are SKA precursor telescopes) data processing, named \textit{Galaxy} and \textit{Magnus} in the Pawsey Supercomputing Centre in Perth, Western Australia. The Pawsey Supercomputing Centre is within 100 meters from the ICRAR-Curtin and CSIRO-CIRA (Perth branch), which provides scientists with easy access to the resources and services of supercomputing facilities, as well as establishes a good academic ecology. 
To better serve the SKA project, the Australian government has approved 70 million Australian dollar (AUD) in 2019 to upgrade the Pawsey supercomputers, which will be able to meet the needs of SKA1. 
The Australian SRC will be established by relying on ICRAR, CSIRO, Pawsey, and universities in eastern Australia, each with a different function. Australia provides a reference scheme 
for the SRC structure, i.e., the SRC to be built in an individual country or region is not a single entity, but a consortium that includes several physical units with different functions (e.g., administrative headquarters, international cooperation, science, computing, storage) together.

Other countries (including the UK, the Netherlands, Spain, Italy, Portugal, Canada, South Africa, etc.) have done a lot of work in the pre-design and preparation stages of the SKA, building and operating the SKA pathfinder projects such as Low Frequency Array (LOFAR), enhanced Multi-Element Radio Linked Interferometer Network (e-MERLIN), etc. Extensive experience has been gained in operating radio interferometric arrays and data processing. Back in 2007, the Netherlands Radio Astronomy Institute (ASTRON) and IBM's institute in Zurich jointly launched the DOME project for the LOFAR data centre \cite{dome}, using the world's most advanced IBM Deep Blue series supercomputers at that time, and carried out high-tech research and development for SKA data analysis. In addition to using the supercomputers in Groningen for data pre-processing, a large amount of data post-processing (equivalent to the tasks of the future SRCs) is transferred to the Dutch national supercomputers in Amsterdam, and the long-term storage of data are placed in three supercomputers in the Netherlands, Germany, and Poland \cite{2022A&A...659A...1S}. The LOFAR observation, data pre-processing, deep data processing, and long-term data archiving actually provide a model of the future SKA Regional Centre. In 2016, European SKA member countries jointly proposed the construction of SKA European regional centre, a project named AENEAS (see also Section 2). The UK government granted the pre-study funding to the UK SKA team, mainly for testing the computing platform prototypes and carrying out the SRCNet prototyping.  The Spanish SRC team built an SRC prototype to support preparatory scientific activities for the future SKA projects, including hardware and cloud computing infrastructure, science archive, software and services, user support and training  \cite{2022JATIS...8a1004G}.
In the next section, we focus on the progress of building and testing the China SRC prototype system.

\section{China SKA Regional Centre} \label{sec:ChinaSRC}

\subsection{Preparation of China SRC} \label{sec:CSRC-prep}

China is the first country in Asia to sign the SKA Observatory Convention and is second only to the UK (hosting the SKA headquarter), Australia and South Africa (hosting the SKA telescopes) in terms of the percentage of investment in the SKA, making China a key member of the SKA Observatory.
China has made important contributions to the SRC preparation (see congratulatory letter from the SKA Headquarter on China's ratification of the SKA Observatory Convention \footnote{\url{https://www.skatelescope.org/news/china-ratifies-skao-convention/}}):  
leading the development of the world's first SKA Regional Centre prototype; jointly completing the largest scale SKA workflow simulation experiment; and using SKA precursor data and the China SRC prototype to train young scientists.

In 2018, the Ministry of Science and Technology (MOST) of China deployed funding to support the prototype construction of the SKA data processing system and scientific pre-research \cite{hong2018ska}. 
In 2021, the MOST launched the SKA special science programs, and SHAO and National Astronomical Observatories of China (NAOC) are leading two high-priority science directions, namely, ``Cosmic Dawn and Cosmic Reionization Detection" and ``Pulsar Search, Timing and Gravity Theory Testing", respectively\footnote{\url{https://service.most.gov.cn/u/cms/static/202009/29081744qk1w.pdf}}. 
The Chinese Academy of Sciences (CAS) 
has funded a key project for international cooperation on the initial construction of the SKA Asia-Pacific Science Data centre. The Science and Technology Commission of Shanghai Municipality has placed the SKA Science centre as one of the key objectives in promoting science and technology innovation.

Under the leadership of the SKA China Office, 
the China SKA team completed  ``Comprehensive Justification Report on China's Participation in the SKA (Phase I)" and the ``China SKA Science Report" \cite{wu2019}, which are the top-level guidelines for China's participation in the SKA. 
From 2016 to 2020, five China SKA Science Conferences have been held with a maximum of more than 200 participants. The SKA science conferences, in conjunction with the SKA special funding, have greatly activated the atmosphere of SKA science research in China and increased the visibility of SKA in the scientific community.
In addition, SHAO held a series of twenty SKA colloquiums in 2021, inviting young scientists from the frontier of SKA research to present the latest SKA research progress to an audience community from domestic institutions and abroad, promoting the academic exchange and growth of young scientists. 

One of the key technical challenges of the SKA is how to properly process the SKA data. This is a common problem faced by SKA science teams worldwide, which will greatly affect the achievement of the expected science goals. SKA is a long-term large science project, now there are 5-10 years to go before SKA1 is completed and fully operational. 
There is plenty of time to train the next-generation SKA science and technology experts. 
For this purpose, several SKA summer schools have been held \footnote{\url{https://indico.skatelescope.org/event/564/}}.  The summer schools combine lectures and practical data processing and use the data obtained from the SKA precursor and pathfinder telescopes to demonstrate the science pipelines and scientific analysis. In such a way, the participants, mostly junior undergraduates, gained an opportunity to acquire a basic understanding of SKA sciences and data processing skills \cite{Contact1}. Similarly, SKAO organizes the SKA Science Data Challenge (SDC). China SRC  was invited to host the second SDC and will also organize SDCs for Chinese users to gain experience in the future global cooperation and operation of SRCs.

\begin{figure*}
    \centering
    \includegraphics[width=0.9\textwidth]{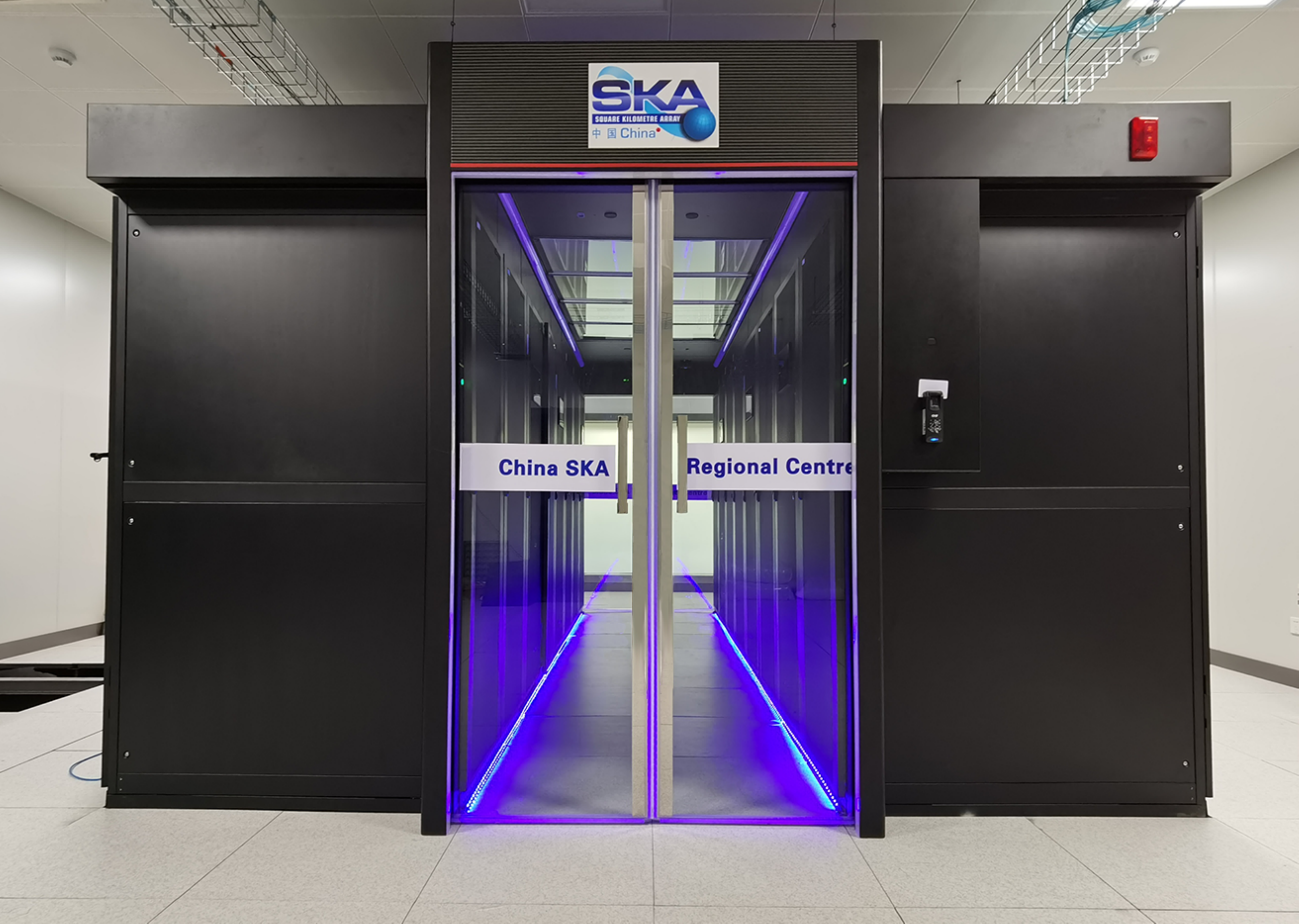}
    \caption{China SKA Regional Centre prototype. The photo was taken in 2020 August. }
    \label{fig:ChinaSRC}
\end{figure*}

\begin{figure*}
    \centering
    \includegraphics[width=0.9\textwidth]{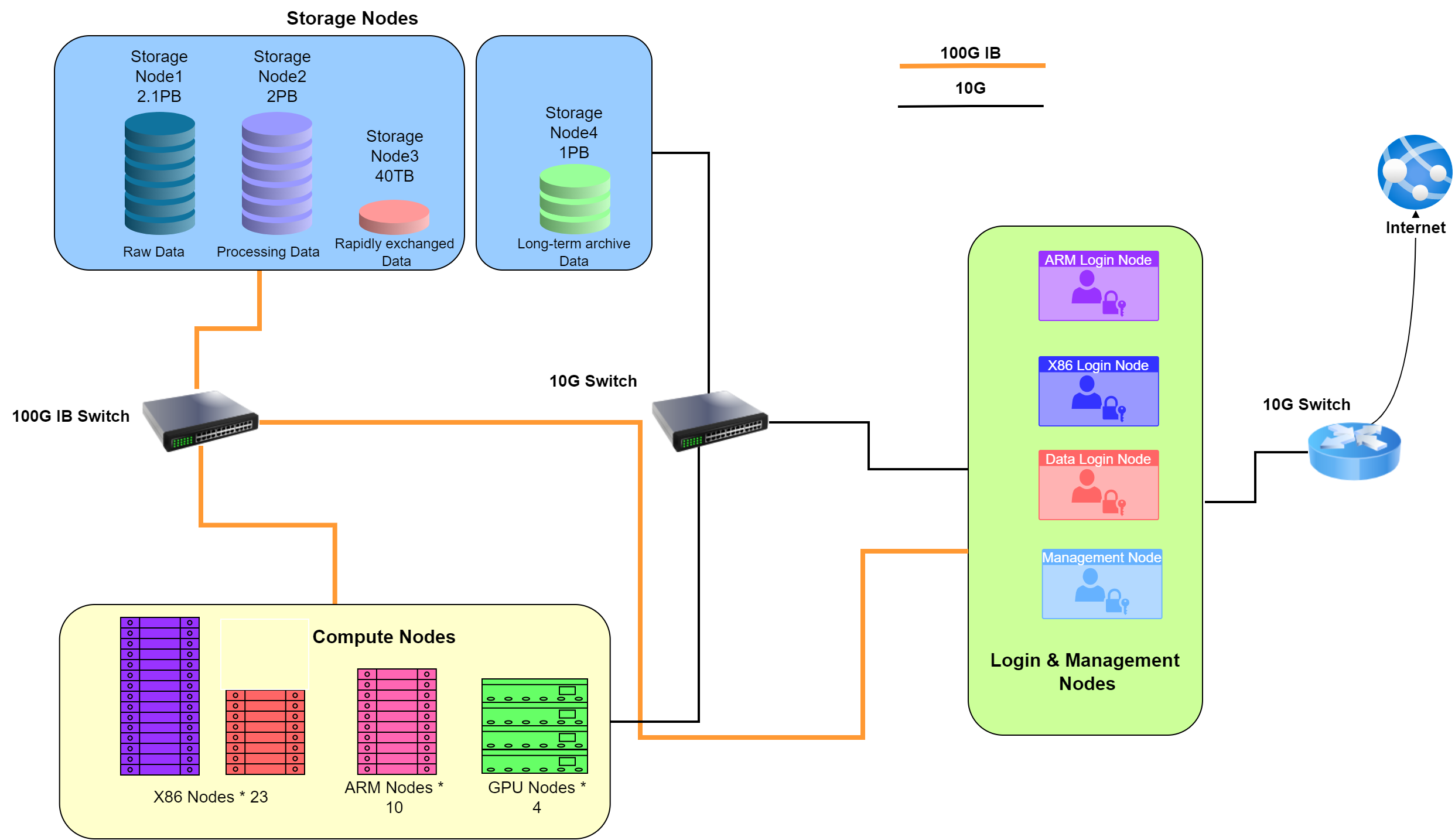}
    \caption{Configuration and topology of the China SKA Regional Centre prototype.}
    \label{fig:ChinaSRCtop}
\end{figure*}

\subsection{China SRC prototype system} \label{sec:CSRCP}

The 6th SKA Engineering Meeting was held in Shanghai from November 25-28, 2019, with nearly 250 participants from all over the world. 
During the conference, the SKA team from SHAO presented the progress of the China SRC prototype. 
SKAO Director-of-General Philip Diamond and SRCSC Chair Peter Quinn co-chaired the advisory meeting of the China SRC prototype. The rapid development and superior performance of the China SRC prototype were highly appreciated by the advisory panel of international experts. 
The evaluation report stated that this is the world's first SRC prototype, and made substantial contributions to advance the global SKA Regional Centre network. 
The successful development of the China SRC prototype 
not only provides a ``Chinese solution" for the global SKA regional centres \cite{SRCWP2020}, but also strongly promotes the scientific operation of the SKA precursor telescopes. 
The China SRC prototype \cite{an2019src} (Figure \ref{fig:ChinaSRC}) has been open to domestic and international users in analyzing and processing SKA precursor and pathfinder data, providing a deeper understanding of the SKA data challenges \cite{2021MNRAS.500.3821B}.
Figure~\ref{fig:ChinaSRCtop} shows the network topology of the China SRC prototype.

\subsubsection{Design} \label{sec:ChinaSRC-design}

The China SRC prototype addresses the characteristics of SKA's data-intensive scientific computing and adopts a design different from that of traditional supercomputers.
In traditional supercomputing centres, the compute and storage networks are separated. This design is suitable for compute-intensive use cases, but becomes very inefficient for I/O-intensive tasks. 
This is reflected in the fact that the traditional supercomputing platform has small local storage, low shared memory capacity, long data invocation time, and single computing architecture. These characteristics are not convenient for the emerging globalized multi-user and highly concurrent, ultra-large dataflow like the SKA. 
SKA has not only very large data volumes, but also large-size individual files, making data movement the biggest bottleneck of the workflow. 
The global shared file system used by traditional supercomputing platforms in their storage architecture often results in high system failure rates for SKA-scale data processing. Moreover,  when using large-scale nodes, many compute nodes are idle when data are not available, causing a great waste of operation cost.

To address the above issues, China SRC adopted the design concept of 'data constellation'. A Data Constellation is a cluster system customized from traditional clusters that integrate compute, storage, and network to bring compute closer to the data and minimize the cost of data movement. A data constellation is a scalable, integrated data processing unit that can be installed on one cabinet or several adjacent cabinets, consisting of a scalable distributed storage system, heterogeneous computing nodes (i.e., X86 Intel CPUs, ARMs, GPUs), and a network system. 
The hybrid heterogeneous computing architecture of the China SRC prototype improves the efficiency of the entire computing system by allocating compute-intensive, memory-intensive and data-intensive tasks of the pipelines to corresponding computing devices in a rational manner, effectively saving the running costs of complex astronomical pipelines and high parallelism challenges.

Each compute node of the data constellation is physically integrated with its very large memory and storage system, which consists of a local storage unit consisting of NVMe solid state drives (SSDs) located at each compute node and a distributed file system consisting of individual storage nodes. Since each data constellation 
can either perform computational tasks individually or form a larger computational system with other data constellations.
Resources can be dynamically and flexibly allocated according to the needs of data processing tasks, either by a single data constellation or by multiple data constellations, and thus can meet a variety of use cases such as multiple pipelines, multiple types of resource requirements, different computational scales and distributed tasks.
In fact, for the current SKA precursor data, we find that most of the data processing tasks can be accomplished within a single data constellation, thus avoiding a large amount of data exchange between independent computing nodes and independent external storage nodes in traditional supercomputing systems.
This data constellation architecture not only can meet the needs of astronomical big data computation and storage, but also can greatly reduce the impact of I/O bottlenecks caused by traditional global file systems.

In practical operation, we found that the SKA precursor workflows are usually memory-bound. Therefore, we deployed large memory on each compute node to alleviate the challenge of processing individual large-sized data files and avoid or reduce the time cost of data cutting, data movement, and idle waits (see Section 3.2.2 for more discussion). Practical experience shows that these extra time costs account for the majority of the runtime of the whole pipeline. In addition, because multiple iterations need to be run in SKA workflows, the large memory capacity enables some files that need to be read repeatedly to reside in memory for long periods of time and be accessed by multiple nodes, thus greatly reducing the time consumption associated with frequent reads in and out of these files and speeding up the workflow.

Another key constraint of data-intensive SKA workflow is the data I/O limitation, which  is a fundamental difference from compute-intensive workflows.
SKA workflows need to handle the storage, curation, acquisition, and processing of large datasets, with most of the processing time consumed in data I/O and the data movement and replication.
Parallel processing of data-intensive computing tasks usually starts by cutting the data into small chunks, each of which can execute the same application independently. This thus requires that the entire data processing system is specifically designed so that the degree of parallelism can scale with the amount of data.
The prototype's multi-tier hybrid storage system contains SDDs and hard disk drives (HDDs) to ensure high performance read and write for a wide range of scientific applications such as high performance computing, high data I/O, and multi-load tasks. In addition, the distributed storage architecture provides a high-throughput, highly concurrent, and highly scalable storage mechanism that ensures near-linear performance growth as the data centre expands to meet the increasing needs of scientific users.
The prototype has a dual-link EDR IB connection between the storage and the compute cluster, solving the most serious data I/O bottleneck in data-intensive computing and ensuring smooth data exchange within and between nodes, reducing not only the latency of dataflow to less than 1 $\mu$sec, about 10 times faster than that of 10~Gbps network but also the risk of system crashes caused by network congestion. Benchmark experiments show that the maximum data flow bandwidth is 2~TB/s for CPU+GPU nodes and 80~GB/s for ARM nodes.

\subsubsection{Overall structure} \label{sec:ChinaSRC-structure}

\begin{table*}[t]
\caption{Hardware system of China SRC Prototype\label{table-csrcp-hardware}}
\begin{center} \bahao  \tabcolsep 5pt
\begin{tabular}{cclccccc}
\toprule
 {Type }          & \#No     & CPU                                                                   &  Memory       & Storage & Cores  \\ \hline
  {X86 Node} &  8         & Intel(R) Xeon(R) Gold 6132 CPU @ 2.6GHz   &    1TB             & 4TB       & 28 \\ 
                      &  15       & Intel(R) Xeon(R) Gold 5218 CPU @ 2.3GHz   &    768GB        & 1.2TB    & 32 \\ \hline
 {ARM Node}&  10         & Kunpeng 920 CPU @ 2.6GHz                        &    1TB             & 600GB   & 96  \\ \hline
  {GPU Node}&  1         & Intel(R) Xeon(R) 2690 @ 2.6GHz   with 4 NVIDIA Tesla V100      &    256GB             & 4TB       & 28 \\ 
                      &  1        & Intel(R) Xeon(R) 6152  @ 2.3GHz   with 4 NVIDIA Tesla V100 (NVLINK)     &    1TB        & 4TB    & 44 \\ 
                       &  1        & Intel(R) Xeon(R) 6140  @ 2.3GHz   with 8 NVIDIA Tesla V100 (NVLINK 32GB)     &    512GB        & 4TB    & 36 \\ 
                       &  1        & Intel(R) Xeon(R) 5320 @ 2.2GHz   with 4 NVIDIA Ampere A40 (40GB)   &    512GB      & 1TB    & 52 \\ \hline
 {Login Node} &  1 & Intel(R) Xeon(R) Gold 6152 CPU @ 2.10GHz @ 2.10GHz  &    754GB   & 1.5TB   & 88  \\ 
  &  1  & Kunpeng 920 CPU @ 2.6GHz  &    64GB   & 1TB   & 64  \\ 
   &  1 & Intel(R) Xeon(R) Gold 6152 CPU @ 2.10GHz  &    692GB   & 250GB   & 88  \\ 
  \hline
\bottomrule
\end{tabular} \\
Columns 4--6 present the specification of each node. 
\end{center}
\end{table*}

The China SKA Regional Centre prototype  currently has 23 Intel X86 CPU nodes, 10 ARM CPU nodes, 4 NVIDIA GPU nodes and 3 login nodes (Table \ref{table-csrcp-hardware}). The execution framework allocates resources to each data constellation and can use either one of the x86 architecture, ARM architecture, and GPU architecture, or a combination of them. The combination can be flexibly customized to accommodate different types of astronomical data (e.g., continuum visibility data, time-domain data, spectral line data) and different computational needs (e.g., data-intensive, compute-intensive, memory-intensive) and different processing requirements for different processing steps/stages in a pipeline (e.g., compute-intensive, data-intensive, memory-bound). Currently, CPU nodes have a total of 1,664 CPU cores with a total 51 TFlops (double precision). The multiple cores up to 96 cores in a single node effectively support highly concurrent tasks. 
The processing of the SKA precursor telescope data reveals that the ARM node has a higher parallel speedup ratio due to its more cores and high memory bandwidth per node than the x86 CPU node when running the same highly parallel multi-process program (see science use case in Section \ref{sec:CSRC-sci-2}).
The GPU nodes consist of sixteen NVIDIA V100 and four A40 GPUs with $\sim$120 TFlops in total. Twelve of these V100 GPUs have NVLINK connectivity of up to 300 Gb/s. 
The GPU nodes are used for both HPC and AI use cases (Sections~\ref{sec:CSRC-sci-2} and \ref{sec:CSRC-sci-5}).
China SRC has 5.1 PB of storage space, of which 55\% of the disk space is already stored with scientific data. To ensure the data exchange rate within the whole system, the storage and computing nodes are all connected via 100G InfiniBand (IB) network.
The prototype is equipped with a high-speed (2--5 Gbps) international network to connect to other SRC nodes \cite{Guo2022-network,Guo2022-simulation}. 

Key radio astronomy programs in data processing need to load data into memory for arithmetic operations, so the capacity of memory determines the performance of the workflow. Moreover, during data processing, a large amount of arithmetic operations take place in memory. However, distributed memory requires a limited number of one-time reads, and when a single data file exceeds a certain size, the computation cannot be completed in a single step. The SKA precursor telescopes' data files are often in the tens of gigabytes or even tens of terabytes, it is obvious that the memory of a single compute node cannot hold the whole data file in some cases. Astronomers have adopted the data-slicing method; that is, the data is first cut into several slices in time sequence, and then each slice is read and processed in turn. Experience shows that the process of cutting the data and loading the data slices into memory consumes a significant amount of runtime, in the scalability tests, the runtime of data loading decreases rapidly by a factor of 3.5--5.8 as the number
of compute nodes increases from 1 to 4, and flattens after 5 \cite{Lao2019w-proj}. The smaller the number of slices, the shorter the overall runtime.
Using distributed memory and deploying the computation to multiple compute nodes for parallel processing do not solve these problems effectively because distributing the computational tasks to multiple compute nodes also needs to  transfer or copy multiple data slices on compute nodes, thus moving data across different compute nodes takes extra time and consumes more time instead. In short, traditional computing platforms have to sacrifice data access time to achieve a balance between computation cost and benefit, which will be quite time-consuming and resource-intensive for data-intensive application scenarios in SRCs.

China SRC uses a design scheme that optimizes the total memory capacity of the compute nodes and the average memory of a single core of the compute nodes, which is used to solve the challenge of processing a single large-sized data file and to avoid or reduce the time cost on data slicing, data movement and idle waiting.
China SRC allocates a large amount of memory to each compute node 
(up to 1~TB memory for CPU nodes, Table~\ref{table-csrcp-hardware}), allowing for processing a data file at once when possible or minimizing the number of data slices. For instance, for a data file with a size of about 100 GB, processing can be done on a single compute node of China SRC prototype, while data files with a size of hundreds of GB only need to be cut into a few 
slices first. The large memory capacity, together with high-speed network and high-performance parallel file system, enables effectively relief the I/O bottleneck in astronomical big data processing.

In addition to the large total memory capacity of computing nodes, the memory scheme also stresses the large amount of memory that can be allocated to individual computing cores (pre-core memory).
Since the processing of large data generated by modern telescopes is essentially accelerated by parallel processing, each thread in the parallelization process is controlled by a single processor core. The amount of data that can be processed by a thread is limited by the amount of memory allocated.
As mentioned above, when running data-intensive tasks, the data slices cannot be too small (otherwise it will increase the total time for data processing), which only sacrifices the advantage of multiple compute cores \cite{Lao2019w-proj}.
Traditional supercomputing platforms primarily serve the computing tasks
and no need to allocate large amounts of memory per thread. 
For a 100 GB file, if it is cut into 25 slices according to time sequence and each slice is assigned to a compute core, only 22 of the total 68 cores can be used. In other words, more than half of the cores are idle and are not used. In contrast, the maximum per-core memory
of the China SRC compute node is 32 GB/core, a factor of 24 higher, allowing the entire file to be read and processed at once.

The bandwidth of data I/O is the third key factor affecting workflow. Memory is the most frequently accessed and interacted component, based on this, China SRC compute nodes place priority on access bandwidth metrics. The total access bandwidth can reach 80 GB/s for ARM nodes, and 2 TB/s for CPU+GPU nodes. 
China SRC connects the hybrid heterogeneous compute node system and distributed storage system through a high-speed network system of up to a total bandwidth of 200 Gbps, and enables internal networking through high-throughput, low-latency IB switches to ensure smooth data exchange within and between nodes, which not only reduces the latency of data flow, but also reduces the risk of system crashes due to network congestion.
The data exchange bandwidth within the compute nodes is directly related to the NVMe SSD Cache bandwidth. Currently, the maximum measured I/O throughput of a single node is 7.4 GB/s, while the theoretical peak is 8.5 GB/s, with an I/O utilization rate of 94\%, which exceeds the conventional standard for HPC cluster. 
In addition, the whole system is directly connected to other SRCs via the  intercontinental network with a total bandwidth of up to 10 Gbps \cite{Guo2022-network,Guo2022-simulation} to support the transmission and management of astronomical big data.

Each data constellation has a corresponding high-performance distributed storage system with hybrid SSD+HDD storage medium. The distributed storage systems of different data constellations are relatively independent and are interconnected through a high-speed, low-latency network system. According to the different requirements for data I/O at different stages of the data processing pipeline, the storage system uses high-capacity memory and SSD local storage units as multi-tier storage.
The storage system is capable of storing, managing, recalling, and reusing data throughout its life cycle, and can meet a variety of applications such as high-performance computing, high data I/O, and multi-load tasks. The file system of the storage system is highly scalable and also provides a redundant backup mechanism that can rebuild data at a rate of 2~TB per hour using redundant backup data.

\subsubsection{Software system}
The China SRC prototype can be regarded as a fully functional mini-supercomputer with complete hardware and software systems. The hardware system is described in detail in Section \ref{sec:ChinaSRC-structure} and the software system is described in detail in this section.
In addition to the software systems necessary for the supercomputer, such as job scheduling system, file system, and management software, the China SRC prototype is also equipped with commonly used radio astronomical data processing software (e.g., CASA~\cite{casa2007}, AIPS~\cite{aips2003}, Difmap~\cite{difmap1997}, Miriad~\cite{miriad1995}) and toolkits (SAOImageDS9~\cite{ds92003}, CARTA~\cite{carta2020}, Topcat~\cite{topcat2005}, etc.). In addition, several scientific data processing pipelines, including the MWA GLEAM/GLEAM-X \cite{2022arXiv220412762H} pipeline, the ASKAP spectral imaging pipeline, the MWA pulsar search pipeline, and the VLBI data processing pipeline, have been established to match the differentiated data processing needs of different scientific directions.
For large-scale scientific data processing tasks, the China SRC team also developed MPI parallel on single/multiple CPU nodes and parallel acceleration on GPU nodes \cite{Lao2019w-proj,2019AcASn..60...12L,Ang2020w-facets,Wei2022presto,Wei2022mwa,Lao2022mwa}, and also developed AI-based data processing software, e.g., deep learning-based source identification and classification software named HeTu \cite{2021SciBu..66.2145L} (Section~\ref{sec:CSRC-sci-5}). 

Astronomical software, including the SKA science pipeline, is constantly evolving and iterating, with varying degrees of dependence on programming tools and versions of the environment. In order to support the software environment needs of different science users, the China SRC prototype has been installed with different versions of compilers, libraries, and basic software for users to easily compile their own programs by simply changing environment variables. During the compilation process, a large number of third-party libraries may be called, as well as complex dependencies between software, which can be realized with the technical support of China SRC.

To facilitate synchronization and sharing software among SRCs, China SRC has also developed a virtual machine and container environment. Containers are also essential elements of the future distributed processing and Cloud-based environment of SRCNet. 
With authorization, users can also create their own containers. China SRC has created containers of MWA, ASKAP and LOFAR data processing pipelines which are greatly helpful for domestic and international users.

\subsubsection{Performance benchmark test}

When the China SRC prototype was initially built in 2019, we used the ASKAP spectral line data and pipeline for performance tests and compared the results with  ASKAP's HPC cluster. 
The $w$-projection imaging algorithm \cite{cornwell2008w-proj} was used in the ASKAP pipeline at that time. The memory and computing cost required in this algorithm are mainly determined by the output image's pixel size ($N_{\rm pix}$), visibility length ($N_{\rm vis}$), field of view ($FoV$) and maximum $w$ ($w_{\rm max}$) \cite{offringa2014wsclean}. When the $FoV$ of the radio telescope is fixed, the pixel size of the output image and the $w_{\rm max}$ will increase exponentially with the increase of the baseline length of the telescope array. Therefore, the memory and computational cost required for the imaging will increase with the baseline length. The experimental results showed that China SRC prototype not only successfully completed the data processing and obtained consistent experimental results with those from the ASKAP HPC cluster, but also exhibited better performance. Specifically, due to the large memory capacity and high-performance storage system of China SRC prototype, it can process the ASKAP data on the 6 km baseline in a single pass. Whereas, the then ASKAP HPC cluster could only process data on 3-km baseline using the same parameters; to execute the complete data processing, the pipeline had to do multiple runs. The results of this comparison were presented during the 2019 SKA Engineering Conference, demonstrating that the design of the China SRC prototype was in line with the technical route of the SKA data processing system.

\begin{figure}[H]
    \centering
    \includegraphics[width=0.5\textwidth]{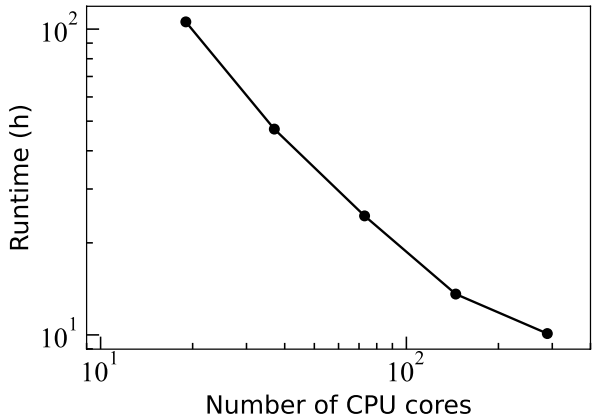}
    \caption{Change of the runtime of a spectral line imaging pipeline with the number of CPU cores used (Ref. \cite{Lao2022mwa}).}
    \label{fig:runtime-askap}
\end{figure}

China SRC has continued to upgrade and improve the configuration and has been able to support data processing for SKA precursor/pathfinder telescopes, e.g. ASKAP and MWA (see science cases in Section~\ref{sec:CSRC-sci}). 
In the MWA GLEAM pipeline, we first performed separate hotspot distribution analysis on ARM and x86 CPU nodes, and then optimized the time-consuming steps. The optimized pipeline improved the overall runtime by a factor of 2.4--2.7, with some programs haveing a speedup ratio of more than 25 times \cite{Wei2022mwa}. It was also found that ARM nodes with more compute cores have higher speedup ratios than x86 CPU nodes. 
For the ASKAP spectral line pipeline, the workflow is typical of distributed processing tasks, a same pipeline can be executed on each frequency channel in parallel. The experiment mainly used seven x86 CPU computing nodes for parallel processing and imaging of data. The pipeline was optimized for MPI execution and deployment, and experiments were conducted using different numbers of CPU cores. The results show that the runtime of the pipeline decreases in an approximate power law (power law index is $-1.1$) as the number of CPU cores increases (as shown in Figure \ref{fig:runtime-askap}), indicating that the parallel process of spectral line imaging is highly scalable \cite{Lao2022mwa}. In addition, we have accelerated the pulsar search pipeline using several parallel optimization methods. The details are given in Section~\ref{sec:CSRC-sci-2}.

The successful deployment of the China SRC prototype is a beneficial solution not only for SKA but also for other data-intensive scientific computing fields. The accumulated operational experience will be useful for the large-scale expansion of SRCs in the future.

\subsection{Construction plan of China SRC}

China SRC undertakes two types of services: SKAO facing and Chinese users facing. The functions and interfaces of these two services are different, see Table~\ref{tab-SRCfunction}.

\begin{table}[H]
\small
\caption{SKAO-facing and Chinese users-facing functionalities\label{tab-SRCfunction}}
\begin{center} \bahao  \tabcolsep 2pt
\begin{tabular}{p{0.23\textwidth}|p{0.23\textwidth}}
\toprule
SKAO-facing functionalities & Chinese users-facing functionalities\\ \hline
A node of the  global SRCNet to ensure stability of data logistics & Primarily channel for Chinese users to access and use SKA data \\
Computing resource provider, especially, for key science projects & ensure resource for key science projects chaired or co-chaired by Chinese astronomers \\ \hline
Long-term storage and curation to maximize the use of SKA science data products & Priority to supporting Chinese scientists in using long-term archive data for resue and data mining, subject to SKA data policies \\ \hline 
Secure authorization and authentication, data access protocols supporting a wide range of access modes & Support Chinese data scientists in the development of big data-related technologies subject to SKA data policies \\ \hline 
Develop and update necessary software environment & Assist users to develop data processing software systems and pipelines \\ \hline
User support, training, documentation & Education, training, and internships for undergraduate and graduate students \\ \hline 
Science popularization and outreach & Science popularization and outreach  \\ 
  \hline
\bottomrule
\end{tabular} \\
\end{center}
\end{table}

The SRCSC has made an assessment of the overall scale of the SRC Network \cite{SRCWP2020}, and the capacity of individual SRCs has not been determined at present. 
If equivalent to China's proportional contribution to SKAO, the baseline capacity of China SRC would need to reach about 2 PFlops computing, about 100 PB/yr storage growth rate, 100 Gbps intercontinental network connectivity, and standardized data processing and management systems. 
The contribution, functions, and responsibilities of China SRC will be confirmed by signing a Memorandum of Understanding with SKAO in the future \cite{SRCWP2020}. 
Some countries would develop more capacity which mainly serves science users in their own countries or regions. The scale of the China SRC is determined by the strategic goal of the China SKA: to play a leading role in radio astronomy in future. This strategic objective is based on China's science and technology capacity, 
and the rapidly growing astronomical community. After considerable discussion among the China SKA community, China SRC is initially planned on the scale of 1/4 of the SRCNet in Phase I, which includes the portion of Chinese contribution to the global SRCNet and the remaining  portion to enhance and strengthen the scientific capabilities of Chinese astronomers. According to this thought, the plan for the China SRC construction is as follows: by 2025, 10 Gbps network, 0.5 PFlops computing capacity and over 20 PB storage; and 100 Gbps network, 5.5 PFlops computing capacity and 1 ExaByte storage by 2030.

China SRC is positioned as a node of the global SRC network and a world-class SKA research center.  
The data center platform of China SRC serves as the main node to interact directly with other international SRCs, and is responsible for the sharing, computing, and management of the Tire-1 data (a total of 700 PB per year). 
China plays a leading role in supercomputing construction in the world, which lays a good foundation for building the data processing infrastructure of China SRC. However, the SRC is not simply a supercomputer; it is more important for its international collaboration, scientific research, and talent training roles, in addition to its international responsibilities for shared computing and storage. 
To maximize SKA scientific output, China SRC will use existing domestic supercomputing and network infrastructures to set up computational subcentres affiliated with major supercomputers, scientific subcentres  where SKA users are grouped, and data subcentres where storage is conveniently accessed.

Based on the experience gained from building and operating the SRC prototypes as well as other data-driven big scientific facilities, the future full-scale China SRC compute platform will be a new type of computing facility (Section \ref{sec:ChinaSRC-design}) that supports data-intensive scientific computing, with high energy-efficiency ratio and high I/O bandwidth:
\begin{itemize}
    \item 
Synchronization with international SRC nodes. The Chinese SRC will be a key node in the global SRCNet, not only responsible for receiving, computing, managing, and archiving Tire-1 data, but also need to exchange data and collaborate with other SRCs. This means that the technical standards, processing software, and other aspects need to be coherent and synchronized. 
    \item 
The SRC computing platform is a dedicated heterogeneous system. Unlike traditional national supercomputers that emphasize node homogeneity to support large-scale operation or provide common computing services, the SKA regional centers need to support numerous scientific working groups with different scientific objectives, widely varying software, and different requirements in different stages of dataflow. Only a hybrid heterogeneous architecture system can effectively perform the SKA data processing tasks.
    \item
Data-centric rather than compute-centric. The positioning of general-purpose supercomputers is to serve the common needs of multiple industry fields, mainly computational task-driven workflows, whereas SKA workflows are typically I/O and memory-intensive workflows.
Huge SKA dataflow puts extremely high demands on I/O bandwidth, CPU memory capacity per core, storage capacity, and storage performance (see Section~\ref{sec:ChinaSRC-structure}). 
    \item 
Long-term service support for science users. The astronomical software keeps evolving, optimization, and upgrades during its lifetime, requiring a stable software development team familiar with radio astronomy and scientific computing to provide decades of continuous maintenance for the SKA data processing software. The development of a large number of SKA scientific applications also requires a dedicated scientific research platform that integrates the joint wisdom of astronomers and ICT (information and communications technology) experts and strengthens internal collaborations. 
    \item
New technologies are constantly being applied to astronomy. For example, artificial intelligence (AI) techniques such as machine learning and deep learning have been applied to classify galaxies and identify pulsars \cite{2021SciBu..66.2145L,2021AcASn..62...20L,Innovation,Xu2022-FRB}. 
SKA data processing requires GPU nodes with large memory capacity and memory bandwidth, very frequent data exchange between GPU nodes, and high parallelism between nodes. 
A computational architecture designed in accordance with the requirements of AI tasks will certainly play an increasingly important role in the SKA science research.
\end{itemize}

The technical challenges and opportunities in building and operating the SKA's end-to-end science system (Eflops-level computing, EB-level storage, 100 Gbps network) will also provide experience for other large facilities planned for the future in the physical domain and across the data-driven scientific and commercial sectors. For example, the next-generation Event Horizon Telescope (EHT) is benefiting from the logistics system development of the SKA.

There are some uncertain factors that may affect the SRC scale and cost\cite{SRCWP2020}: 
\begin{enumerate}
\item 
    Execution efficiency of radio astronomy computational methods and algorithms. It is closely related to the computational architecture, system complexity, and degree of parallelization of specific pipelines, and can directly affect the scale and cost of the computing nodes. 
    The typical processing efficiency of SKA pathfinders' workflows is expected to be $\sim$11\%, which is also used for estimating the computation requirements of SRCs and SDPs \cite{Scaife}, 
    but still needs to be practically validated in SKA early science data and also tested in state-of-the-art distributed HPC/cloud computing systems.
\item  
    Data storage volume. The SRC will receive over 700 PB data per year around 2028, so the largest cost of the future SRC infrastructure will come from data storage and management.  The scheme 
    of data backups will determine the total amount of data stored in SRCNet, and data averaging and compression techniques will also have a significant overall impact on the final storage costs.
\item 
    The proportion of Cloud used in SRC. Both the SRC Working Groups and the SRCNet prototyping groups are investigating the practicability of a hybrid model of HPC and Cloud architecture for SRCs (Figure~\ref{fig:SRCwg}). If a significant amount of data processing is transferred to Cloud, the capital costs of HPC hardware acquisition and upgrade, and the operational costs of supercomputing centers, can be translated into a single cost of purchasing Cloud resources that are elastic and scalable based on demand.
\item 
    Unequal levels of national infrastructure participation in the SRC Network. The reality of infrastructure varies from country to country, and the SRCSC allows and encourages multiple forms of input from the national or regional stakeholders. The degree of openness and policies (free or charged) of different levels and sizes of research infrastructure and educational research networks will have a significant impact on the SRC cost budget.
\end{enumerate}

\section{Science use cases of China SRC} \label{sec:CSRC-sci}

The China SRC prototype has successfully deployed the SKA precursor/pathfinder telescopes' data processing pipelines (including containerization of the pipelines), stored about 1.5 PB of scientific data, and established a high-speed transcontinental broadband network of up to 5 Gbps connecting with other SRC nodes (e.g., Australian SRC). 
Some representative scientific use cases are as follows.

\subsection{Simulation and parameter inference of global EoR signals} \label{sec:CSRC-sci-1}

Probing the cosmic dawn and epoch of reionization (EoR) is the highest priority scientific direction for the Chinese SKA science strategy \cite{wu2019}. Observing the redshifted neutral hydrogen 21 cm signal in the low-frequency radio band will provide key information to answer frontier scientific questions in cosmology, such as what are the sources of cosmic reionization and how the ratio of neutral hydrogen evolves during the reionization. 
However, detecting the HI 21 cm signal from the cosmic dawn and EoR is extremely difficult. 
In the case of the global 21 cm signal detection, how to tightly constrain the physical parameters and how to separate the artificial spectral structures from the underlying cosmic signal are key issues that should be addressed before a successful detection can be performed.

The detection of the EoR signal requires a long period of accumulation of observational data. It is necessary to simulate the observation mode and to evaluate the parameters before carrying out the final observation. 
Dr. Junhua Gu and Dr. Jingying Wang performed a simulation of the global HI 21 cm signal of the cosmic dawn and EoR, from which to inspect what information one can extract from the global HI 21 cm signal, from which they examined what information can be extracted from the global HI 21 cm signal in future real observations and the key factors affecting the real observations \cite{2020MNRAS.492.4080G}.
This simulation includes a large-scale statistical parameter inference based on a parallel computing algorithm that can be significantly accelerated by a large number of usable CPU cores. 
Each ARM node in China SRC prototype has 96 cores (see Section \ref{sec:CSRCP}), and there are 10 ARM nodes. If all ARM compute nodes are used, then up to 960 compute cores can be used, which is very helpful for a compute-intensive use case like this one.

The instrumental effects of the digital beamforming system of the SKA-Low 
may affect the scientific results. Therefore, having this information in advance through simulations will help to study the future processing of real data. In real observations, the real-time data processing is independent for each individual antenna in SKA-Low stations, so the workflow can be naturally accelerated by multiprocessing parallelization. The simulation experiments of the digital beamforming system of SKA-Low have been completed using China SRC.

The common feature of the above two use cases is that the computational tasks are composed of a set of almost independent subtasks and can therefore be deployed to all available CPU cores. 
As little data is exchanged among subtasks, the  acceleration in such cases is approximately proportional to the number of CPU cores available and is extremely suitable for supercomputing clusters with a large number of CPU cores. Light-weight computing tasks can also benefit from a high number of cores on a single node. In both cases, the early versions of the programs were coded with serial algorithms. By adding very few directives to the source codes, the program can be immediately parallelized on a single node and gain significant performance acceleration.
J.-H. Gu and J.-Y. Wang (Ref. \cite{2020MNRAS.492.4080G}) conducted a series of large-scale Monte Carlo-Markov sampling computations. For each individual problem, about eight million reionization signal spectra were involved in the computation. The wall-clock time consumed was about 72 hours using two ARM nodes of the CSRCP, each using 84 CPU cores. The powerful parallel computing capability provided by the China SRC prototype makes this simulation feasible, and the use of multi-node parallelism will be considered in the future to further increase the computational speed.

\subsection{Pulsar search}\label{sec:CSRC-sci-2}

Pulsar is one of the two highest-priority science directions of the SKA in China. Astronomers have been searching for pulsars using the China SRC prototype \cite{2020SSPMA..50j9501G,2021AcASn..62...20L}. The most commonly used pulsar software is PRESTO\footnote{\url{https://www.cv.nrao.edu/~sransom/presto/}}. However, the PRESTO-based pulsar search pipeline is very slow when dealing with large samples of data. Optimization methods for the pulsar search pipeline are mostly based on GPU and FPGA acceleration. Whereas most astronomical software is still developed based on CPU architecture, therefore it is necessary to use multi-core CPU clusters to accelerate the pipeline.

The China SRC prototype is equipped with both X86 and ARM CPU clusters, both with multi-node and multi-core configurations. We have accelerated the pulsar search pipeline based on PRESTO \cite{Wei2022presto}, using OpenMP parallelism and load balancing, along with Shell script multiprocess, respectively. From the standard pulsar search workflow, we found five steps ``hot spots" which are time consuming and can be efficiently optimized on our machine. These five steps are ``Remove Narrowband RFI", ``De-dispersion", ``Accelsearch", ``Candidate Folding" and ``Single Pulse Search". We applied OpenMP parallelism for RFI removal, multi-core parallelism for de-dispersion, and load balancing for the rest three hot spots (see details in \cite{Wei2022presto}). Our strategy is to pursue optimal solutions for parallelism, especially for more than 3000 independent time series data based on different fine channels.

The final optimization was tested with observational data obtained from the SKA-Low precursor MWA, using its four incoherently summed data with a time resolution of 0.1 ms, and observation times spanning from 10 -- 80 mins (3072 fine channels). The results show that the optimized pipeline runs more than 10 times faster on the 28-core X86 node and more than 20 times faster on the 96-core ARM node than the pipeline using one core on two types of nodes without optimization. Since ARM has more CPU cores than X86, the pipeline on the ARM cluster runs faster than that on the X86 cluster by a factor of 1.1--1.3, showing the potential of the ARM clusters used in the SKA data processing.
This performance optimization is not only applicable to time-consuming processes such as pulsar searches but can also be extended to other science pipelines to accelerate astronomy applications on the China SRC.

\subsection{Discovery of an ultra-long period magnetar} \label{sec:CSRC-sci-3}

Transient objects such as supernova explosions and gamma-ray bursts light up the sky, but normally, the low-frequency radio sky has been very quiet. However, the discovery of a low-frequency transient by cutting-edge SKA precursor telescopes is unveiling a new era of radio transient astronomy \cite{2022Natur.601..526H}.

An international research team led by Dr. Hurley-Walker from ICRAR discovered an abnormal radio transient source with long-period low-frequency emission by analyzing observational data made with the MWA. The source has not been detected in subsequent optical, infrared, or high-energy band observations. Dispersion measurements of its radio pulse indicate that this transient source should be located in the Milky Way at a distance of about 1.3 kpc
from the Sun. Polarization measurements show that the linear polarization of the transient source is about 90\%, which exceeds all known pulsars in the same observational mode in the 150 MHz band, indicating the presence of super strong magnetic fields. The long period and high polarization in the low-frequency band of the source cannot be explained by the theoretical models and observational features of known pulsars. 
It could be an ultra-long-period magnetar or a white dwarf with super-strong magnetic fields.

The discovery of this transient source was made possible by the high-sensitivity SKA low-frequency precursor telescope and the dedicated computing clusters tailored to the characteristics of the SKA data. The search for this class of objects employs a model-subtracting time-step imaging technique that requires searching for notable residuals in all voxels. This technique is ten times more sensitive than the fast search usually employed, but consumes enormous computational resources \cite{2022Natur.601..526H}. 
These observations generated a huge amount of raw data, and an even larger amount of data are generated by intermediate processes. For example, in polarimetric imaging, the number of image files produced exceeds 10 million, an order of magnitude higher than the number of Stokes I images. The large volume of datasets and great complexity of the data processing process give the data processing software extremely high challenges on the data input/output (I/O) bandwidth, high concurrent tasks, and highly parallelized processing of the computing cluster. 
In this project, China SRC performed very well in the processing of the wide-band polarization data and some polarization image analysis, as well as shared the storage of the MWA data. 

This discovery is the first detection of a long-period transient source in the Galactic plane region, and opens a new window for the search of low-frequency transient sources \cite{2022Natur.601..526H}. If more transient sources with similar characteristics can be found, it will help to fully understand the evolution and death of stars and fill the gap in the study of magnetars.

\subsection{Simulating SKA1-scale workflow} \label{sec:CSRC-sci-4}

The SHAO SKA team, in collaboration with research teams from ICRAR (Australia) and ORNL (US), completed a full-scale, complete-cycle simulation experiment simulating the SKA1 workflow on the world's fastest supercomputer at that time \cite{2020SciBu..65..337W,2020hpcn.conf...11W}. 
The experiment successfully simulated an ``end-to-end" SKA1 workflow, including a real-time pipeline of observation data generation, collection, correlation, calibration, and imaging, with more success than expected. It demonstrates the pipeline processing of a typical SKA science experiment. It simulated a 6-hour observation of the SKA1-Low and the entire process ran continuously for 3 hours with an average computing speed of 65 PFlops and a peak of 130 PFlops, reaching the design specifications of SKA1 supercomputer (the SDP). 
More than 27,000 GPUs and nearly 200,000 CPU cores were used in data processing, and the total data volume reached 2.6 PB. The maximum data write rate was as high as 925 GB/s,  which is 1.5 times the design specification of the SDP and more than 3 orders of magnitude higher than the conventional workflow, indicating that astronomers have a preliminary solution to the scientific computing problem of massive  I/O load similar to SKA.

The joint team also developed the dataflow execution framework (named DALiuGE) specifically for the SKA \cite{2017A&C....20....1W}. 
\footnote{Because Chinese scientists made significant contributions in the early development of DALiuGE, the `Liu' in the name adopts the phonetic translation of the Chinese word 'flow'.} 
DALiuGE adopts the advanced idea of data-driven processing, which not only greatly improves the scalability, but also improves the operational efficiency, and has good flexibility and adaptability in fault tolerance \cite{2017A&C....20....1W}. The intelligent resource scheduling method of DALiuGE can significantly reduce the power consumption of the overall workflow, which has practical significance for large-scale scientific computing. DALiuGE has been used for data processing of radio telescopes (for example, the survey programs of the ASKAP) and scientific applications of the SRC \cite{2019ASPC..523..183K,2022AJ....163...59D,2020ASPC..527..531G}. 
DALiuGE not only meets the data processing requirements of SKA, but also has the potential value to be applied in the field of data-intensive scientific computing. 

The early validation of the scalability of DALiuGE was led by the SHAO team on the Tianhe-2 supercomputer and gained tremendous attention in the HPC field and public media.
After the completion of China SRC prototype, some key technology verification  experiments were carried out on China SRC prototype. During the 2019 SKA Engineering Meeting, Tao An, Andreas Wicenec, and Rodrigo Tobar made a live demonstration experiment on behalf of the joint team, which attracted great attention of the attendees.
 
The success of this SKA1-scale workflow experiment has set a milestone for the largest data stream in astronomy and increased the confidence of astronomers in building SRCs \cite{Contact2}. The key technological breakthrough achieved in this experiment has strong applications in other large-scale scientific computing fields, and was therefore selected as one of the six finalists for the 2020 Gordon Bell Prize in High Performance Computing.

\subsection{Artificial intelligence for source finding}
\label{sec:CSRC-sci-5}

Artificial intelligence, one of the most cutting-edge ICT technologies, is being widely applied to astronomy. The astronomers at the Shanghai Astronomical Observatory have developed a source-finding software using artificial intelligence and named it ``\textsc{HeTu}" after a traditional Chinese mythological story \cite{2021SciBu..66.2145L}. 
\textsc{HeTu} aims to provide an automated and intelligent data processing method for classifying galaxies in large-scale radio astronomical image sets.

\textsc{HeTu} is based on convolutional neural networks and performs best in image processing. Specifically, \textsc{HeTu} uses a combined network structure with a residual network (ResNet) and a feature pyramid network (FPN) as the backbone network. It combines the advantages of both: the residual network can balance the recognition accuracy and computational cost; the pyramid network can detect objects of different scales in the same picture. Therefore, \textsc{HeTu} not only improves the accuracy, but also provides the feature map of multi-scale objects, which is exactly what astronomers need. 
In the testing experiment, the average precisions for the compact source class and double-lobe source class are 0.994 and 0.981, respectively, which are better than other software for similar purposes.

\textsc{HeTu} has been used for source detection and classification based on images observed by the SKA-Low precursor MWA \cite{2017MNRAS.464.1146H}. \textsc{HeTu}'s detection results were compared with those obtained by \textsc{Aegean} \cite{2018PASA...35...11H}, a source-finding software using the traditional component-based fitting method. 
The cross-match rate between the \textsc{HeTu}-detected and \textsc{Aegean}-detected point-like sources is 96.9\%, indicating the practical potential of \textsc{HeTu}.
For extended sources, component-based fitting methods fit the source as multiple Gaussian components, and cannot determine the connection between adjacent components, rendering that the identification and classification of extended sources cannot be done automatically. \textsc{HeTu} has a natural advantage in identifying extended sources based on source morphology for recognition and classification. 
\textsc{HeTu} found 100\% FRII-type radio galaxies and 97.4\% FRI-type galaxies. 
In addition, the running speed of \textsc{HeTu} is 21 times faster than that of \textsc{Aegean}, and two orders of magnitude faster than visual recognition.

Automated and accurate source-finding tools are particularly important for image analysis of ongoing and upcoming large sky surveys, and the recognition performance and speed of \textsc{HeTu} will be further improved in the future to support larger-scale image processing and to focus more on peculiar morphological objects, such as odd radio circles \cite{2021PASA...38....3N,2022MNRAS.513.1300N}. The vastness of the Universe provides the neural network with a naturally large amount of data for training (learning), and on the other hand, the development of artificial intelligence will undoubtedly have a profound impact on astronomy.

\subsection{Imaging the fine structure of compact objects with VLBI} \label{sec:CSRC-sci-6}

VLBI is the highest resolution observing technique, and the recently Black Hole Imaging made by the Event Horizon Telescope is a perfect demonstration of VLBI's unique observing capability. 
VLBI is one of the science directions of SKA  and one of the main observing modes of SKA \cite{2015aska.confE.143P}.
SKA1-Mid, SKA1-Low, and FAST can join the global VLBI network as a giant telescope element, which can improve the sensitivity of the global VLBI network by one order of magnitude. 
Moreover, SKA2 will extend the baseline from $<$100 km of the SKA1 to 3000--5000 km, so the fully completed SKA is a super VLBI network itself! 
The scientific preparation of SKA-VLBI, including the organization, observation model and supporting software, data processing software, and science case design, has been initiated. \footnote{\url{https://whova.com/web/vlbis_202111/}}.

China has an independently operated Chinese VLBI network \cite{2018NatAs...2..118A}, which plays an important role in high-precision quasi-real-time VLBI orbit determination in lunar exploration missions and deep space exploration missions. China has the largest 500-m radio telescope, FAST, which is expected to join the low-frequency VLBI in the future. 
Currently, China SRC has established a pipeline for VLBI data processing \cite{Lao2022mwa}, which facilitates VLBI data processing for SKA users. 
Representative scientific results using this pipeline include: 
(a) VLBI observations of a number of active galactic nuclei in the early Universe have yielded high-resolution images that reveal the compact structure and kinematics of the jets, their kinematic ages, and interstellar environments \cite{2021MNRAS.507.3736Z,2021MNRAS.506.1609C,2020NatCo..11..143A,2020ApJS..247...57C,2020SciBu..65..267A,2020SciBu..65..525Z,2022A&A...662L...2Z,2022A&A...663A.139A,2022MNRAS.511.4572A}. All these critical information improves the understanding of the nature of the radio emission from the first-generation supermassive black holes and jets feedback to their host galaxies. 
(b) High-resolution observations of transient sources were carried out with VLBI to help understand their physical nature. After the discovery of the ultra-bright optical transient AT2018cow, VLBI follow-up observations, spanning over one year period, revealed compact radio afterglow emission \cite{2020ApJ...888L..24M} and captured the afterglow evolution in the transition phase and later ($\sim$90 days and beyond).
The absence of collimated relativistic jets rules out the possibility of a tidal disruption event or ordinary core-collapse supernovae. The progenitor may be a relatively low-mass star that exploded to produce a magnetar. 
(c) very high-energy TeV gamma-ray bursts are very rare events, and their energy release mechanisms are of great value for understanding high-energy astrophysics. 
High-quality VLBI images of a TeV GRB 190829A were obtained from the EVN+eMERLIN array and VLBA observations, on a total of 9 epochs over a period of 117 days \cite{2021arXiv210607169S}. 
The observations set tight limits on the size and expansion rate of the radio source, which are in good agreement with the source size evolution obtained by detailed modeling of the multi-band light curves with the forward+inverse shock model. The VLBI data break the parameter degeneracy that often occurs in afterglow modelling, and offer a successful test of the standard gamma-ray burst afterglow model. 

\section{Conclusion and Perspective}

In the coming months, the SRC community will be conducting global coordinated test of the SRCNet prototype systems. 
China SRC will be in the first batch of nodes in the global SRC network. At the same time, the conceptual study of the construction and organization of the domestic SRC sub-network will be initiated.

In the coming years, the China SRC team will first complete the construction of about 10\% of the scale of the China SRC (i.e. Phase 1) following the standardized management and technical standards of the global SRC network. In parallel, the layout and initial construction of the domestic SRC network will also initiated.
The organization and structure of the China SRC will fully consider the geographical distribution of SKA astronomical research centers, computing centers and data storage centers with different functions, and optimize the allocation of resources.

2026-2028 is the critical time period for the full-scale construction of the China SRC and its integration into the global SRCNet. By then, the China SRC will enable to support the key science projects (KSPs) and Principal Investigator (PI)-led projects. 
In the long-term operation phase after full completion, China SRC aims to become a world-class research platform for SKA data processing, storage, curation and international cooperation; it will also be an outstanding SKA research center to which scientists around the world aspire.

Rapidly developing Chinese ICT industry has a solid technical foundation to support the SKA Regional Centres. The design capacity, production capacity and product performance of general-purpose processor chips, accelerator chips, interconnect chips and NPU chips are approaching the first class in the world. 
Storage in terms of tiered architecture, multiple redundancy methods, reconstruction techniques, parallel file system is also rapidly developing, with order of magnitude performance improvement in IOPS (Input/output operations per second), latency, etc. Artificial intelligence (AI) as a general-purpose technology is being gradually applied to various science areas \cite{Innovation} and is gradually spreading in astronomy \cite{2021SciBu..66.2145L}. China SRC will not only help Chinese astronomers to get priority access to advanced SKA data product, but also greatly enhance the innovation capability in supercomputing and big data processing technologies.

China's scientific and technological development needs international cooperation, and the open international science project also needs China's participation.
Astronomers around the world are engaged to exploring scientific questions of common concern to mankind under the framework of international cooperation in the SKA Regional Centres.

\section*{Abbreviations}

\begin{footnotesize}
\begin{verbatim}
ALMA - Atacama Large Millimeter/submillimeter Array
ASKAP - Australian Square Kilometre Array Pathfinder
CAS - Chinese Academy of Sciences
China SRC - China SKA Regional Centre
CSRC-P - China SKA Regional Centre prototype
CSIRO - Commonwealth Scientific and Industrial Research 
Organisation
CSRCP - China SKA Regional Centre Prototype
e-MERLIN - enhanced Multi-Element Radio Linked Interferometer Network 
EVN - European VLBI Network
GLEAM - GaLactic and Extragalactic All-sky MWA
GLEAM-X - GaLactic and Extragalactic All-Sky MWA-eXtended
HPC - high performance computing
HST - Hubble Space Telescope
ICRAR - International Centre for Radio Astronomy Research 
LOFAR - Low Frequency Array
MOST - Ministry of Science and Technology
MWA - Murchison Widefield Array
NAOC - National Astronomical Observatories of China
NSFC - National Natural Science Foundation of China
ORNL - Oak Ridge National Laboratory
RFI - radio frequency interference
SAFe - Scaled Agile Framework
SDC - SKA Data Challenge 
SHAO - Shanghai Astronomical Observatory
SKA - Square Kilometre Array 
SKA1 - the first phase of the SKA
SKA2 - the second phase of the SKA
SKA-Low - the low-frequency telescopes of the SKA
SKA-Mid - the mid-frequency telescopes of the SKA
SKAO - Square Kilometre Array Observatory
SRC - SKA Regional Centre
SRCNet - SKA Regional Centre Network
SRCSC - SKA Regional Centre Steering Committee
VLT - Very Large Telescope
VLBA - Very Long Baseline Array
VLBI - Very Long Baseline Interferometry
ICT- Information and Communication Technology
\end{verbatim}
\end{footnotesize}

\Acknowledgements{
This work was supported by National Key R$\&$D Program of China (2018YFA0404603), Chinese Academy of Sciences International Partner Program (114231KYSB20170003), National Natural Science Foundation of China (12041301) and Youth Innovation Promotion Association, Chinese Academy of Sciences (201664, 2021258). TA thanks William Garnier, Junhua Gu and Jianwen Wei for helpful discussion and comments on the manuscript. }




\end{multicols}
\end{document}